\documentclass[manuscript]{aastex}

\usepackage{graphicx,times}

\shorttitle{Magnetar Giant Flare Model}
\shortauthors{Meng et al.}

\begin{document}

\title{An MHD Model For Magnetar Giant Flares}
\author{ Y. Meng\altaffilmark{1, 2}, J. Lin\altaffilmark{1}, L. Zhang\altaffilmark{3}, K. K. Reeves\altaffilmark{4},
 Q.S. Zhang\altaffilmark{1, 5}, and F. Yuan\altaffilmark{6}}
\email{mengy@ynao.ac.cn,jlin@ynao.ac.cn}
\altaffiltext{1}{Yunnan Observatory, Chinese Academy of Sciences, P. O. Box 110, Kunming, Yunnan 650011, China.}
\altaffiltext{2}{University of Chinese Academy of Sciences, Beijing 100039, China.}
\altaffiltext{3}{Department of Physics, Yunnan University, Kunming, Yunnan 650091, China}
\altaffiltext{4}{Harvard-Smithsonian Center for Astrophysics, 60 Garden Street, Cambridge, MA 02138, USA}
\altaffiltext{5}{Key Laboratory for the Structure and Evolution of Celestial Objects, Yunnan Astronomical Observatory,
Chinese Academy of Sciences, Kunming, Yunnan 650011, China}
\altaffiltext{6}{Key Laboratory for Research in Galaxies and Cosmology, Shanghai Astronomical Observatory, Chinese Academy of Sciences,
80 Nandan Road, Shanghai 200030, China}

\begin{abstract}
Giant flares on soft gamma-ray repeaters that are thought to take place
on magnetars release enormous energy in a short time interval. Their power
can be explained by catastrophic instabilities occurring in the
magnetic field configuration and the subsequent magnetic reconnection.
By analogy with the coronal mass ejection (CME) events on the Sun, we
develop a theoretical model via an analytic approach for magnetar giant flares.
In this model, the rotation and/or displacement of the crust
causes the field to twist and deform, leading to flux rope formation in the
magnetosphere and energy accumulation in the
related configuration. When the energy and helicity stored in the configuration
reach a threshold, the system loses its equilibrium,
the flux rope is ejected outward in a catastrophic way, and magnetic reconnection
helps the catastrophe develop to a plausible eruption.
By taking SGR 1806 - 20 as an example, we calculate
the free magnetic energy released in such an eruptive process and find that
it is more than $10^{47}$ ergs, which is enough
to power a giant flare. The released free magnetic energy is converted into
radiative energy, kinetic energy and gravitational
energy of the flux rope. We calculated the light curves of the eruptive processes
for the giant flares of SGR 1806 - 20, SGR 0526-66 and SGR 1900+14,
and compared them with the observational data.
The calculated light curves are in good agreement with the observed light curves
of giant flares.

\end{abstract}


\section{Introduction}
Soft gamma repeaters (SGRs) and anomalous X-ray pulsars (AXPs) are believed to
be magnetars - a small class of spinning neutron stars with ultra-strong
magnetic fields ($\emph{B}$ $\geq$ $10^{15}$ G ),
which are thought to result from dynamo action during supernova
collapse \citep{DT92,Kel98,TM01,Lyu03,HL06}. The emission from a magnetar is powered by the
dissipation of non-potential (current-carrying) magnetic fields in the
magnetosphere \citep{DT92,TD96,TLK02,Lyu06}. Both SGRs and AXPs show
quiescent persistent X-ray and repeated soft gamma-ray
emissions \citep{Mel04}. Extremely rarely, an SGR produces a giant flare with
enormous energy ($\backsimeq 10^{44}-10^{47}$ erg) and long burst duration.
These exceptionally powerful outbursts begin with a very short ($\sim$ 0.2~s)
spike of $\gamma$-rays containing most of the flare energy and the
spike is followed by a pulsating tail lasting a few hundreds of
seconds \citep{Hur05}.

So far, three SGRs have been reported to produce giant flares \citep{Maz79,Hur99,Hur05}.
They include SGR 0526-66 on 5 March 1979 \citep{Maz79}, SGR 1900+ 14 on
27 August 1998 \citep{Hur99,Kou99,Vrb00}, and SGR 1806-20 on 27 December
2004 \citep{Hur05,Pal05}. The giant flare from SGR 1806- 20 was much
more luminous than the other two events \citep{Hur05,Pal05}.
Its initial $\gamma$-ray spike released an
energy of $\sim 10^{46}$ erg within $\sim 0.2$~s, and its rising and
falling times were $\tau_{rise}\leq 1$ ms and $\tau_{fall}\approx 65$ ms,
respectively. The main spike was followed by a tail with $\sim 50$ pulsations of
high-amplitude at the rotation period (7.56s) of SGR 1806 - 20 \citep{Hur05,Pal05}.

Although the energy of a magnetar outburst is widely believed to
come from the star's magnetic field, details of the physical process
in which the magnetic energy is stored and released remain
unknown. So far, two models of giant flares of SGRs exist, which
depend on the location where the magnetic energy is stored prior to the
eruption: one assumes that the energy is stored in the crust of the
neutron star (crust model) and the other one assumes that the storage
occurs in the magnetosphere (magnetosphere model).
In the crust model, a giant flare is caused by a
sudden untwisting of the internal magnetic field
\citep{TD95,TD01,Lyu06}. Subsequently, a large and quick rotational
displacement on the time-scale of a flare leads to the giant flare.
Alternatively, in the magnetosphere model, the magnetic energy is slowly
stored in the magnetosphere on time scales
much longer than that of the giant flare itself, until the
system reaches a critical state at which the equilibrium becomes unstable.
Then further evolution in the system occurs and leads to
flares, in analogy with solar flares and coronal mass ejections (CMEs)
taking place in the solar atmosphere \citep{Lyu06}. Observations of the
giant flare from SGR 1806-20 on 27 December 2004 showed that it lasted a
very short rise time, $\sim$ 0.25~ms \citep{Pal05}. This time interval
of the eruption is short compared to
the time-scale required for the crust model
\citep{TD95,Lyu03,Lyu06}. Therefore, at least for the SGR 1806 - 20,
the time-scale of the crust model is too long to account for the
triggering and early stage evolution of the event.

The magnetosphere model based on an analogy with solar
CMEs was proposed by \citet{Lyu06}. In this model, the magnetic
energy released during a giant flare is built up slowly in the
magnetosphere, not in the crust of the neutron star. Although the
energy storage process is gradual and long, the energy release takes
place within a very short timescale in a dynamical fashion \citep{Lyu06}.
Transition from slow to fast evolution constitutes the
catastrophe that is similar to what happens in solar flares and CMEs \citep[see][]
{FI91,Iea93}.
 Moreover, the profile of the light
curve of solar-flare/ CME events resembles that of magnetar
giant flares. Both of them have an impulsive phase and a tail
emission. Similar morphology and characteristics between
 magnetar giant flares and solar-flare/ CME events indicate
the operation of a common physical mechanism.
Therefore, solar flares and CMEs give an important
prototype context for magnetar giant flares. \citet{Mas10}
constructed a theoretical model for a magnetar giant flares based
on the solar-flare/ CMEs model. They described magnetar giant
flare using a magnetic reconnection model of a solar flare proposed
by \citet{SY99} while taking account of chromospheric
evaporation. In their work, the preflare activity produces a
baryon-rich prominence. Then the prominence erupts as a result of
magnetic reconnection, and the eruption constitutes
 the origin of the observed radio-emitting ejecta
associated with the giant flare from SGR 1806- 20
\citep{Tay05,Cam05,Gae05,Mas10}. A giant flare should be induced
as the final outcome of prominence eruption accompanied by large-scale
field reconfigurations \citep{Mas10}.

Numerical simulations have also been performed in order to construct an
MHD model for a magnetar giant flare.
\citet{Pfr12a,Pfr12b,Pfr13} presented the simulations of evolving strongly
twisted magnetic fields in the magnetar magnetosphere. Their results showed that slow
shearing of the magnetar crust leads to a series of magnetospheric expansion
and reconnection events, corresponding to X-ray flares and bursts. They studied the
relationship between the increasing twist and the spindown rate of the star, and
concluded that the observed giant flares could be caused by the sudden opening
of large amounts of overtwisted magnetic flux, resulting in an abrupt
increase in spin period.

Motivated by CME studies, \citet{Yu11} constructed a  general
relativistic model of non-rotating magentars, which simulated
how the magnetic field could possess enough energy to overcome
the Aly-Sturrock energy constraint and open up. Furthermore, by
taking into account the possible flux injections and crust motions,
\citet{Yu12} built a force-free
magnetosphere model with a flux rope suspended in the magnetosphere
and investigated the catastrophic behavior of the flux rope in a background
with multi-polar magnetic field. In this model,
a gradual process leads to a sudden release of the magnetosphere
energy on a dynamical timescale \citep{Yu12}.
Therefore, the existing catastrophe model for solar eruptions
could be a good template for constructing a theoretical model
for magnetar giant flares.

The initiation and development of magnetar giant flares have been
extensively studied and several models have been suggested \citep{TD95,TD01,Lyu06,
Mas10,Yu11,Yu12}. But the origin and development of these giant
flares remains unclear. The detailed physical process of the
magnetic energy storage and release is still an open question.
In this work, we consider relativistic effects and construct a
magnetohydrodynamical (MHD) model for magnetar giant flares in the
framework of the CME catastrophe model \citep{LF00}, then duplicate
the dynamical process of the giant flares produced by SGR 1806- 20.
We describe our model in next section. Results of
calculations and comparisons with observations are in Section 3.
Finally, we discuss these results and summarize this work
in Section 4.

\section{Model Description}
In the framework of the catastrophe model of solar eruptions, the evolution in the system
that eventually leads to a plausible eruption includes two stages and
a triggering process that initiates the second stage. In the first stage,
the magnetic energy is gradually
stored in the coronal magnetic field in response to a motion or
change inside the star, and the stored energy is quickly released in
the second stage. The second stage follows the first one via a
triggering process, which is also known as the loss of the equilibrium.
This process could be either ideal or non-ideal MHD depending on the fashion
of the system evolution or detailed structure in the magnetic field involved
\citep[e.g., see discussions of][]{FI91,Lel03}.
But magnetic reconnection is crucial for the energy  released to produce
the flare \citep{Par63,Pet64} and to allow the CME to propagate
smoothly \citep{LF00}.

\begin{figure}[fig1a]
\epsscale{1}
\plotone{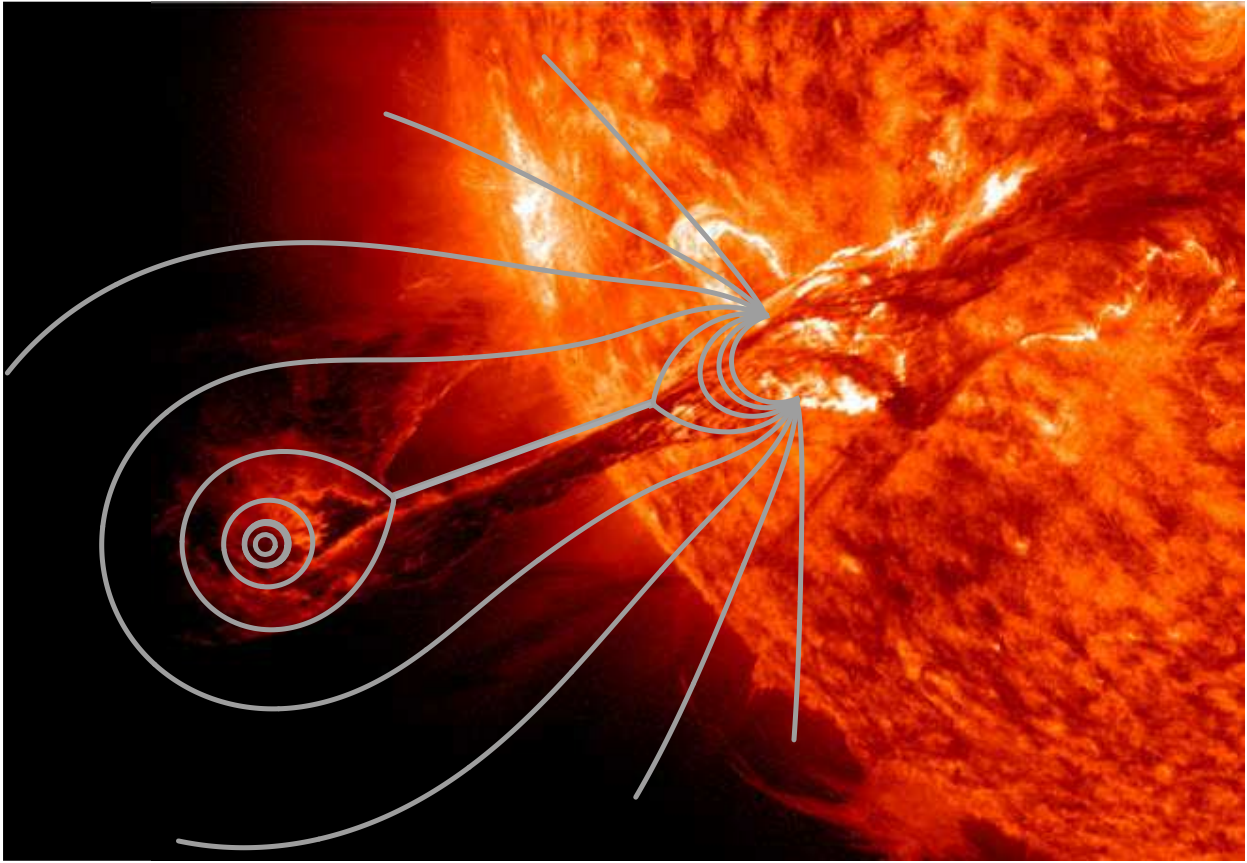}
\caption{Example of a typical eruptive event (namely a CME) occurring in the solar atmosphere
(the original image was taken from the SDO website: http://sdo.gsfc.nasa.gov/),
overlapped with a sketch for describing the surrounding magnetic field in a plane
perpendicular to the main axis of the ejecta.}\label{Fig.1a}
\end{figure}

Figure \ref{Fig.1a} shows a typical eruptive event or process (or CME) in the solar magnetic atmosphere
(the original figure was obtained from the SDO website: http://sdo.gsfc.nasa.gov/).
In this process, because of the instability, the magnetic configuration
including a large amount of high temperature plasma is ejected outward from the solar
surface, associated with intensive electromagnetic radiation (namely the well-known solar
flare) and energetic particles \citep[e.g., see also][]{Sve76,PF02}.
Since the highly ionized plasma and the magnetic field are frozen to one another \citep{Pre82},
the plasma is confined in the nearby magnetic configuration. Therefore, we are able to
infer magnetic structure and some internal details according to the level of
concentration or spatial distributions of the ambient plasma, which is shown as
different bright or dark features in the picture displayed in Figure \ref{Fig.1a}, although
we cannot see the magnetic structure itself directly.

The complex global and local structures in the magnetic configuration shown
in Figure \ref{Fig.1a} indicate high non-potentiality of the associated magnetic field
and strong interactions between the magnetic field and the electric current in that
region. Hence, as a result, enough magnetic free energy (i.e., the difference
between the total magnetic energy in a system and the magnetic potential energy
in the same system) prior to the eruption can be stored to drive the eruption.
A sketch for qualitatively describing the disrupting magnetic field is inserted
in the figure, with more details being specified in Figure \ref{Fig.1}: the closed circles
in the middle represents the core of the CME, the ambient curves represent the
magnetic field lines associated with the CME. More explanations for mathematical
notes will be given later. Since it is surrounded by strong magnetic fields and plasma, a
magnetar is quite likely to possess similar complex magnetic configurations in
its magnetosphere and eventually to produce energetic eruptions, which must be,
of course, much more powerful than those occurring on the Sun.

We note that the driver of the energy transport occurring in the solar
atmosphere is the motion of the dense plasma in the photosphere, but the crust
of the magnetar is solid and might not be able to move as the mass in the
photosphere. However, crust cracking could take place on the magnetar
from time to time \citep[see][]{Rud91a,Rud91b,Rud91c}.
The surface magnetic field of a spin-down crust-cracking neutron star might break up
into large surface patches (platelets) which would move apart from one another
\citep{Rud91c}. On the other hand, the Lorentz force due to the strong
magnetic field acting on the crust could result in the build up of stress, and
further causes pieces of the broken crust to rotate on the equipotential surface as the
stress acting on the lattice exceeds a critical value \citep{Lyu06}. This processes
eventually leads to the storage of magnetic stress and thus energy in the magnetosphere.

Therefore, the energy driving the giant flare on the magnetar could be transported
from the inside of the magnetar via a reasonable mechanism, and then stored in the
magnetosphere before the eruption, although the basic driver of energy storage is
somewhat different from that on the Sun. More effects of the footpoint motions of
the magnetic field on the magnetar were also studied by \citet{TD95}.
They studied the effect of a sudden
shift in footpionts of the magnetospheric field (see Figure 1.(a) in their paper)
on the energy release. The effect of crust motions on the release of magnetic energy
was also studied in the numerical simulations by \citet{Yu12}.

\citet{LF00} developed an analytic model of solar eruptions
(Figure \ref{Fig.1}) that explains the mutual impact of
magnetic reconnection and the CME acceleration on each other. Magnetic
reconnection plays very important role in the eruption. It helps to eject the
flux rope successfully and to form a CME;
it produces flare loops, flare ribbons and the rapid
expanding CME bubbles \citep{Lel04,LS04}; and its
non-ideal MHD properties lead to a natural avoidance of the Aly-Sturrock
paradox \citep{Aly91,Stu91},which was first noticed by \citet{Aly84},
such that a purely ideal MHD process in the force-free
environment could not fully open the closed magnetic field
to produce CMEs \citep{For00,Kli01,Lel03}.

\subsection{Equations For Basic Magnetic Configuration}
In this work, we build an  MHD model for magnetar giant
flares based on a solar flare/CME model, including the effects of
special relativity. Since the magnetars that have so far been observed
have a very slow rotation, and the time scale of the main spike of
the magnetar giant flare is short compared to their rotation periods,
we ignore rotation effects on the eruption in our model.
But we understand that the rotation should play an important role
in the first stage.
We now elaborate the possible magnetic configuration that includes
a flux rope floating in the magnetosphere and may give an eruption
eventually. The precursor of a giant
flare could be closely connected to the pre-existence of the
flux rope that includes twisted magnetic field \citep{Got07,GH10,Yu12}.

We note here that our model is plane symmetric instead of axial-symmetric,
and the evolutionary behavior of the disrupted magnetic field
revealed in this work yields valuable and important observational consequences. But we
understand as well that our calculation is performed in two-dimensions. The cartoon in
Figure \ref{Fig.1} could be considered as the cross section of a three dimensional
configuration, which includes a flux rope with two ends anchored to the crust of the
central star such as those shown in Figure\ref{Fig.1a}. Therefore, the effect of the
anchorage of both the flux rope ends and the expelling force due to the curvature of
the flux rope \citep[see also][]{Ltl98,Ltl02,IF07} in reality are not included in
our calculations. Furthermore, the impact
of the centrifugal force due to the spin of the neutron star on the evolution of the
disrupting magnetic configuration is not included, either.

Generally, the anchorage tends to prevent the outward motion of the flux rope, and
the other two forces play the opposite role in governing the evolution of the system.
The expelling and attracting forces acting on the flux rope might counterbalance one
another and show no net effect on the system evolution eventually. But details need
to be studied carefully, which would constitute the main content of our work in the
future in a more realistic environment.

According to the widely accepted physical scenario of the magnetar
\citep{DT92,Kel98,TM01,Lyu03,HL06}, the magnetosphere of a slowly
rotating neutron star includes a strong magnetic field, and
the magnetic field is rooted in the crust of the star. The rotation
and/or the displacement of the crust cause the field to twist and
deform leading to flux rope formation and energy accumulation
in the magnetosphere.

Gravity and magnetic forces in this configuration is acting on
the flux rope \citep{LMV06}. The gas pressure is much smaller than the
magnetic forces, so it is negligible in this model. Prior to the loss of
equilibrium, the evolution in the system is ideal and magnetic reconnection
does not take place in the magnetosphere; and the gradual evolution in the
system in response to the slowly varying boundary conditions at the magnetar
surface eventually causes the loss of equilibrium in the configuration to
occur in a catastrophic fashion.

 During this process, the magnetic energy is
accumulating slowly until it reaches a critical value as shown in Figure \ref{Fig.2},
where the flux rope is at a critical position, and the catastrophic loss of equilibrium
occurs and the flux rope is thrust outward. Following the loss of equilibrium,
the flux rope is thrust outward and its motion is governed by
\begin{eqnarray}
  m \gamma^{3} \frac{d^{2}h}{d t^{2}} = \frac{1}{c}|\textbf{\emph{I}}
 \times\textbf{\emph{B}}_{ext}|- F_{g}
 \label{Fmg}
\end{eqnarray}%
to the first order of approximation,
where \emph{m} is the total mass inside the flux rope per unit length,
$\gamma=1/ \sqrt{1-\upsilon^{2}/c^{2}}$ is the Lorentz factor,
\emph{h} is the height of the flux rope from the magnetar surface,
$\textbf{\emph{I}}$ is the total electric current intensity flowing
inside the flux rope, $\textbf{\emph{B}}_{ext}$ is the total external
magnetic field measured at the center of the flux rope, and $F_{g}$
is the gravitational force acting on the mass inside the flux rope
(see Appendix for details).

In zeroth-order approximation, the following equations
hold \citep[see][for details]{LF00}:
\begin{eqnarray}
  \textbf{\emph{j}} \times \textbf{\emph{B}}=0,
  \label{jbE}
\end{eqnarray}%
\begin{eqnarray}
  \textbf{\emph{j}}=\frac{c}{4\pi} \nabla \times \textbf{\emph{B}},
  \label{dbE}
\end{eqnarray}%
where ${\bf j}$ and ${\bf B}$ are the electric current density and
the magnetic field in the system, respectively. In the Cartesian
coordinate system $(x, y)$, the $x$-axis is on the star surface and
$y$-axis points upward (see also Figure \ref{Fig.1}). Solving
equations (\ref{jbE}) and (\ref{dbE}) gives the description of the
force-free magnetic field in the system \citep{RF05}
\begin{eqnarray}
  B(\zeta)=\frac{2 i A_{0}\lambda(h^{2}+\lambda^{2})\sqrt{(\zeta^{2}+p^{2})(\zeta^{2}+q^{2})} }
  {\pi (\zeta^{2}- \lambda^{2})(\zeta^{2}+h^{2})\sqrt{(\lambda^{2}+p^{2})(\lambda^{2}+q^{2})}},
  \label{Bxy}
\end{eqnarray}%
where $\zeta=x+iy$, $A_{0}= B_{0}\pi \lambda_{0}$ is the source field strength and
$B_{0}= 2 I_{0}/ ( c \lambda_{0} ) \sim 10^{15}$~G is the  magnetic field
strength on the surface of the magnetar.
The corresponding vector potential function $\emph{A}(\zeta)$ is as follows:
\begin{eqnarray}
 A(\zeta)\label{Axy}&=&  - \int {B(\zeta)} d\zeta \nonumber \\
  &=& \frac{{2i{A_0}\lambda {p^2}}}{{\pi q\sqrt {\left( {{\lambda ^{^2}} + {p^2}} \right)\left( {{\lambda ^{^2}} + {q^2}} \right)} }}
\left\{ \left(1 + \frac{{{q^2}}}{{{\lambda ^2}}}\right)\Pi \left[ {\tan^{-1}}\left(\frac{\zeta}{p}\right),1 + \frac{{{p^2}}}{{{\lambda ^2}}},\frac{{\sqrt {{q^2} - {p^2}} }}{q} \right] \right. \nonumber \\
 &+ &\left. \left(\frac{{{q^2}}}{{{h^2}}} - 1\right)\Pi \left[{\tan^{-1}}\left(\frac{\zeta}{p}\right),1 - \frac{{{p^2}}}{{{h^2}}},\frac{{\sqrt {{q^2} - {p^2}} }}{q}\right] \right\},
\end{eqnarray}
where $\Pi$ is the incomplete elliptic integral of the third kind.
According to \citet{LF00}, the current in the flux rope is:
\begin{eqnarray}
 I=\frac{c \lambda A_{0}}{2 \pi h}\frac{\sqrt{(h^{2}-p^{2})(h^{2}-q^{2})}}{\sqrt{(\lambda^{2}+p^{2})(\lambda^{2}+q^{2})}}.
\end{eqnarray}%

\subsection{Energetics}
The law of the energy conservation determines the total energy in the system at a
given time that reads as \citep[see also][]{Rev06}
\begin{eqnarray}
 W_{mag}+W_{KE}+W_{EM}+W_{gra}=W_{0},
  \label{wtal}
\end{eqnarray}%
where $W_{mag}$, $W_{EM}$, $W_{KE}$, $W_{gra}$ and $W_{0}$ are the free magnetic energy,
the radiative energy, the kinetic energy of the flux rope, the gravitational
potential energy and the initial total energy (a constant) in the system, respectively.
Then, we have
\begin{eqnarray}
 \frac{d}{dt}(W_{mag}+W_{KE}+W_{EM}+W_{gra})= 0.
  \label{w0}
\end{eqnarray}%
From Equation (\ref{Fmg}), we find that the power related to the kinetic energy is given by
\begin{eqnarray}
\label{dwk}\frac{{d{W_{KE}}}}{{dt}}&=&\frac{{d\left[ {{m_0}(\gamma  - 1){c^2}} \right]}}{{dt}} \nonumber \\
&=&{m_0} \gamma^{3} \dot{h}\frac{d^{2}h}{d t^{2}}\nonumber \\
&=&\frac{{I{B_{ext}}\dot h}}{c} - \frac{{G{M_{NS}}\gamma{m_0}}}{{{{\left( {{R_0} + h} \right)}^2}}} \dot{h} \nonumber \\
&=&\frac{{{B_0}^2{\lambda ^4}}}{{8h{L_{PQ}}^2}} \left[\frac{{{H_{PQ}}^2}}{{2{h^2}}} - \frac{{({p^2} + {\lambda ^2})({h^2}
- {q^2}) + ({q^2} + {\lambda ^2})({h^2} - {p^2})}}{{{h^2} + {\lambda ^2}}}\right]\dot{h}\nonumber \\
&-&\frac{{G{M_{NS}}\gamma{m_0}}}{{{{\left( {{R_0} + h} \right)}^2}}} \dot{h},
\end{eqnarray}%
where $H_{PQ}=\sqrt{(h^{2}-p^{2})(h^{2}-q^{2})}$, $L_{PQ}=\sqrt{(\lambda^{2}+p^{2})(\lambda^{2}+q^{2})}$,
 $G = 6.67 \times 10^{8}$~cm$^{3}$~g$^{-1}$~s$^{-2}$  is the gravitational constant,
$M_{NS}\sim {M_ \odot } \thickapprox 1.989 \times 10^{33}$ g is the mass of neutron star,
${m_0}= 10^{20}$~g~cm$^{-1}$ is the total mass per unit length inside the flux rope.
And $q$, $p$ and $h$ are functions of time.

The impact of the magnetic field
on the motion of the flux rope is epitomized by the magnetic compression [the first
term in the square bracket at the right hand side of the last equation in (9)] and
the magnetic tension [the second term in the square bracket at the right hand side
of the last equation in (\ref{dwk})]. Generally, the magnetic compression results from the
magnetic field lines between the magnetar surface and the flux rope, which tends to
push the flux rope away from the central star; and the magnetic tension is produced
by the magnetic field lines overlying the flux rope with two ends anchored to the
surface of the central star \citep[see also Figure 10 of][]{Lel03}. The magnetic
field in the solar corona, in fact, dominates the plasma as well, and it plays the
same role in driving the solar eruption as in the case of the magnetar giant flare.
The last term in the last equation of (\ref{dwk}) comes from the gravity of the mass in the
ejecta. The loss of equilibrium in the system takes place as the balance among
these forces ceases to exist, and the flux rope is thrust outward with its motion
being governed by equation (\ref{Fmg}) or (\ref{dwk}). The relativistic effect on the motion of the
flux rope in this process is indicated by the Lorentz factor appearing in the
relevant equations.

The power associated with the radiative energy is given by \citep{Rev06}
\begin{eqnarray}
 \label{st} \frac{dW_{EM}}{dt}&=&S(t) \nonumber \\
 &=&\frac{c}{2 \pi} E_{z}(t) \int^{q(t)}_{p(t)}B_{y}(0,y,t)dy,
\end{eqnarray}%
where $E_{z}(t)$ is the electric field in the reconnection region induced in the reconnection
process \citep[see details given by][]{FL00, LF00}, the magnetic field
along the current sheet $B_{y}(0,y,t)$ is determined by equation (\ref{Bxy}) with
 $x=0$, $q$ and $p$ are the top and the bottom tips of the current sheet, respectively, as shown
in Figure \ref{Fig.1}. Here the product of the electric field $E_{z}$ and the magnetic field $B_{y}$
gives the Poynting flux that describes the electromagnetic energy flux entering the current sheet
with the reconnection inflow.

In this work, we follow the practice of \citet{RF05} using $S(t)$ to
represent the output power of the radiative energy. Because most of the released
magnetic energy is converted into thermal energy in a solar eruption, \citet{Rev06}
only consider the thermal energy instead of radiative energy .
We are able to do so because $S(t)$ is a kind
of description for the amount of magnetic energy brought into the current sheet per unit time
by the reconnection inflow, and it is eventually dissipated by reconnection and converted
into radiative and kinetic energy of the reconnected plasma, as well as the kinetic energy of
energetic particles accelerated, in the current sheet. The radiative energy could account for both the
thermal and non-thermal components of the emission observed, and the energetic particles may produce non-thermal
emission through various ways, such as synchrotron, cyclotron, bremsstrahlung, and so on.
Therefore, it is the conversion of this part of magnetic energy that could contribute to the
emission accounting for the observed light curve in the eruption \citep[see also discussions by][]{RF05}.

Observationally, the consequences of the energy conversion in the eruption are two-fold.
First of all, like its solar counterpart, the current sheet itself is a dense heat source
because a reasonably large amount of the magnetic energy is converted into
heat there as indicated by observations and models of the solar eruptive cases
\citep[e.g., see][and references therein]{Rev10,Qiu12,Cia13,Liu13,Sus13}.
Since the magnetic field is stronger at low altitudes than at high altitudes, most of the
energy released by reconnection mainly occurs in the early stage of the eruption when the
ejecta is low and the sheet is short, and the hottest part of the current sheet is then
 close to the star surface \citep[see e.g. Figure 9 of][]{Rev10}.
 Second, in addition to the direct heating inside the
current sheet, a large amount of energetic particles and a heat conduction front are also
created by reconnection, and then propagate downward along magnetic field lines
\citep[e.g., see][and references therein]{Rev07,Win11,Zha11,Liy13}.
They eventually reach the star surface, yielding
observational consequences.

The kinetic energy of particles is quickly converted into thermal
and non-thermal energy due to collisions of particles with the star
surface, and the conduction front dumps all the thermal energy it brings to
the surface region leading to the further heating. Therefore, a very hot region, namely
the ``fireball", would be naturally expected to form near the star during the giant flare
process \citep[see also Figure 4d of ][]{Mas10}. [A schematic description of this
scenario can be found in Figure 1 of \citet{Lel05}, which describes the origin of the
solar counterpart of the ''fireball", and interested readers are referred to that work.]
Thus, the trapped ''fireball" as discussed by many authors
\citep[e.g., see also][for example]{TD95,Hur05} is a straightforward and natural
consequence of our model, and the emission that comes from this area may account for
various observational consequences.

From Faraday's Law, $E_{z}(t)$ is given by
\begin{eqnarray}
\label{Ez}E_{z}(t)= -\frac{1}{c}\frac{\partial A_{0}^{0}}{\partial t}
= M_{A}V_{A}B_{y}(0,y_{0})/c,
\end{eqnarray}%
where $A_{0}^{0} =A(0,p\leqslant y \leqslant q)$ is the magnitude of the vector potential
along the current sheet, $M_{A}$ is the Alfv\'{e}n Mach number of the reconnection inflow
and is a measure of the reconnection rate in the current sheet \citep{LF00},
is directly proportional to $E_{z}(t)$, and equals to the reconnection inflow
speed compared to the local Alfv\'{e}n speed near the reconnection region.
In this work, it is taken to be a constant at $y_{0}=(p+q)/2$, the height of the current
sheet center, and the magnetic field
$B_{y}$ is evaluated at $y_{0}$. In the inner magnetosphere of magnetar, the Alfv\'{e}n
velocity $V_{A}\simeq c$ \citep{Lyu06}.

By calculating the Poynting flux in the current sheet, we can obtain
the power related to the energy dissipated in current sheet.
Substituting equations (\ref{Bxy}) and (\ref{Ez}) into equation (\ref{st})
and integrating, we have
\begin{eqnarray}
S\left( t \right)&=&- \frac{c}{{2\pi }}{\left( {\frac{{2{I_0}}}{c}} \right)^2}\frac{{{M_A}{\lambda ^2}{{({h^2}\label{Sst}
  + {\lambda ^2})}^2}\sqrt {({y_0}^2 - {p^2})({q^2} - {y_0}^2)} }}{{q({h^2} - {y_0}^2)({y_0}^2 + {\lambda ^2})({p^2}
 + {\lambda ^2})({q^2} + {\lambda ^2})}}  \nonumber \\
   &\times & \left[\frac{{{p^2} + {\lambda ^2}}}{{{h^2} + {\lambda ^2}}}\Pi \left(\frac{{{q^2} - {p^2}}}{{{q^2} + {\lambda ^2}}},
  \frac{{\sqrt {{q^2} - {p^2}} }}{q}\right) + \frac{{{h^2} - {p^2}}}{{{h^2} + {\lambda ^2}}}\Pi \left(\frac{{{q^2} - {p^2}}}{{{q^2} - {h^2}}},
  \frac{{\sqrt {{q^2} - {p^2}} }}{q}\right) \right. \nonumber \\
   &-& \left. K(\frac{{\sqrt {{q^2} - {p^2}} }}{q})\right],
\end{eqnarray}
where $K$ and $\Pi$ are the complete elliptic integral of the first kind and the third kind,
respectively.

According to the law of the universal gravitation, the rate of change in the gravitational
potential energy is
\begin{eqnarray}
\frac{dW_{gra}}{dt}= \frac{{G{M_{NS} }\gamma{m_0}}}{{{{\left( {{R_0} + h} \right)}^2}}} \dot{h}.
\label{dwg}
\end{eqnarray}
Combining equations (\ref{w0}), (\ref{dwk}), (\ref{st}) and (\ref{dwg}), we get
\begin{eqnarray}
\frac{dW_{mag}}{dt}= -\frac{{{B_0}^2{\lambda ^4}}}{{8h{L_{PQ}}^2}}
\left[\frac{{{H_{PQ}}^2}}{{2{h^2}}} - \frac{{({p^2} + {\lambda ^2})({h^2}
- {q^2}) + ({q^2} + {\lambda ^2})({h^2} - {p^2})}}{{{h^2} + {\lambda ^2}}}\right]\dot{h} - S(t),
\label{dwm}
\end{eqnarray}%
which is the expression of Poynting's theorem for the system studied in the
present work \citep[see also][]{RF05,Rev06}.

We define the time when the neutral point forms as $t_{n}$ and $W_{EM}(t_{n})=0$
because no magnetic reconnection occurs at this time, and the magnetic energy at $t_{n}$
is \citep{FP95}
\begin{eqnarray}
W_{mag}(h_{n},t_{n})= (\frac{A_{0}}{\pi})^{2} \left[J_{n}^{2} \ln \left(\frac{2 h_{n}J_{n} }{r_{00}} \right)
+ \frac{1}{2}J_{n}^{2}+\frac{1}{4} \right],
\end{eqnarray}
where $J_{n}$ and $h_{n}$ are correspondingly the current and the
height of the flux rope at this time, respectively,
$r_{00}$ is the radius of the flux rope at the maximum
current point.

According to \citet{Lel04} and \citet{LS04}, ejecta brings a large
amount of magnetic flux away from the star as well. This flux includes
two components: one is associated with the flux rope before the eruption,
and another one is brought from the background to the ejecta by magnetic
reconnection through the current sheet. So the former is constant during
the eruption, and the latter increases after reconnection commences causing
the ejecta to expand rapidly. Freezing of the magnetic field to the plasma
yields that the magnetic flux and the associated plasma sent to the ejecta
by reconnection constitute the outer shell of the ejecta, and the flux is
given by
\begin{eqnarray}
\Phi_{B} = L\left[\frac{I_{0}\pi}{c}-A(0,q)\right],
\end{eqnarray}%
where $A(0,q)$ is the value of Equation (\ref{Axy}) at the upper tip
of the current sheet, which is \citep{LF00}
\begin{eqnarray}
A(0,q) = \frac{{2{I_0}}}{c}\frac{\lambda}{q L_{PQ}}\left[(h^{2}-q^{2})K\left(\frac{p}{q}\right)(q^{2}-p^{2})
+ \Pi \left(\frac{\lambda^{2}+p^{2}}{\lambda^{2}+q^{2}},\frac{p}{q} \right)
-\frac{H_{PQ}^{2}}{h^{2}}\Pi \left(\frac{p^{2}}{h^{2}},\frac{p}{q} \right) \right] .
\end{eqnarray}
Given $p$, $q$, $h$, and $\dot{h}$,
we can obtain the total magnetic flux brought into interstellar space by
magnetic reconnection as a function of time $t$ in the whole process of
the eruption.

\subsection{Equations Governing Motions Of Ejecta}
As the system evolves dynamically and the ejecta moves upward at speed $\dot{h}$,
the electric field $E_{z}$ in equation (\ref{Ez}) can be written as
\begin{eqnarray}
E_{z}(t)&=&-\frac{1}{c}\frac{\partial A_{0}^{0}}{\partial t}\label{dA00}
= -\frac{1}{c}\frac{\partial A_{0}^{0}}{\partial h}\dot{h}\nonumber\\
&=&-\frac{\dot{h}}{c}\left(\frac{\partial A_{0}^{0}}{\partial p}p'+ \frac{\partial A_{0}^{0}}
{\partial q}q'+ \frac{\partial A_{0}^{0}}{\partial h} \right)\nonumber\\
&=&-\frac{\dot{h}}{c}(A_{0p}p'+ A_{0q}q'+ A_{0h}),
\end{eqnarray}
where $\dot{h}= dh/dt$, $p'=dp/dh$ and $q'=dq/dh$.
To find the variations of $\dot{h}$, $p$ and $q$ versus the flux rope
height $h$, we need two other equations: the frozen-flux condition at the surface
of the flux rope and the dynamic equation describing the flux rope acceleration
due to the force acting upon it.

For the frozen-flux condition at the surface of the flux rope, we get
\begin{eqnarray}
\frac{2 I_{0}}{c}A_{R}= A(0, h- r_{0})= \textrm{const}\label{ffc},
\end{eqnarray}%
where $r_{0}$ is the radius of the flux rope. Taking the total derivate about $h$ on
the both side of the equation (\ref{ffc}) gives
\begin{eqnarray}
\frac{\partial A_{R}}{\partial p}p'+ \frac{\partial A_{R}}{\partial q}q'\label{dAR}
+\frac{\partial A_{R}}{\partial h}
= A_{Rp}p'+ A_{Rq}q'+ A_{Rh} = 0.
\end{eqnarray}%
Here $p'$ and $q'$ can be obtained from (\ref{dA00}) and (\ref{dAR}):
\begin{eqnarray}
p'=\frac{\tilde{A_{0h}}A_{Rq}- A_{Rh}A_{0q}}{A_{Rp}A_{0q}- A_{0p}A_{Rp}},
\label{dpdh}
\end{eqnarray}%
\begin{eqnarray}
q'=\frac{A_{Rh}A_{0p}- \tilde{A_{0h}}A_{Rp}}{A_{Rp}A_{0q}- A_{0p}A_{Rp}},
\label{dqdh}
\end{eqnarray}%
where $\tilde{A_{0h}}= c E_{z}/ \dot{h}+ A_{0h}$ and the electric field $E_{z}$
is determined by equation (\ref{Ez}).

The dynamic equation is taken from equation (\ref{dwk})
 \begin{eqnarray}
\gamma^{3}\frac{d^{2}h}{d t^{2}}=\frac{{{B_0}^2{\lambda ^4}}}{{8h{m_0}{L_{PQ}}^2}}
\left[\frac{{{H_{PQ}}^2}}{{2{h^2}}} - \frac{{({p^2} + {\lambda ^2})({h^2}
- {q^2}) + ({q^2} + {\lambda ^2})({h^2} - {p^2})}}{{{h^2} + {\lambda ^2}}}\right]
-\frac{{G{M_{NS} }\gamma}}{{{{\left( {{R_0} + h} \right)}^2}}}.
\label{dh2dt}
\end{eqnarray}%
Eventually, we obtain equations that govern the motion of the flux rope after
the catastrophe:
\begin{eqnarray}
 \label{dht}\frac{dh}{dt}&=& \dot{h},\\
 \frac{dp}{dt} &=& p'\dot{h},\\\label{dpt}
\frac{dq}{dt}&=& q'\dot{h},\\ \label{dqt}
\frac{d^{2}h}{d t^{2}}&=&\frac{{{B_0}^2{\lambda ^4}}}{{8h{m_0}\gamma^{3}{L_{PQ}}^2}}
\left[\frac{{{H_{PQ}}^2}}{{2{h^2}}}
 - \frac{{({p^2} + {\lambda ^2})({h^2}\label{d2ht2}
- {q^2}) + ({q^2} + {\lambda ^2})({h^2} - {p^2})}}{{{h^2} + {\lambda ^2}}}\right]\\ \nonumber
&-&\frac{{G{M_{NS} }}}{{{{\gamma^{2}\left( {{R_0} + h} \right)}^2}}}.
\end{eqnarray}%

So far, we are ready to investigate the dynamical properties of the system following the
catastrophe by solving differential equations (\ref{dht}) through (\ref{d2ht2}). We also
further study the energetics of the system via solving equations (\ref{st}), (\ref{dwg}),
and (\ref{dwm}) consequently by considering equations (\ref{Sst}). We note here that equations (\ref{dpdh})
and (\ref{dqdh}) should be used when we solve equations (\ref{dht}) through (\ref{d2ht2}).

\section{Results}
The characteristic values of several important parameters that are specifically for
the SGR 1806 - 20 and the associated environment around it can be given according to
the observational events we collected so far. These parameters are $R_{0}$, the radius
of the neutron star, $\lambda_{0}$, the half-distance between the two point-sources as
the catastrophe occurs (see Figure \ref{Fig.1} ), $m$, the mass initially included in
the flux rope, and $L$, the length of the flux rope.

We note here that the equations listed previously for governing the emission or light curve
of the giant flare and the dynamic properties of the associated mass ejection look very complex
and include many parameters, but only a couple of them are free
parameters and the others can be fixed for the specific cases. These parameters include the
mass of the neutral star $M_{NS}$, the initial scale of the eruption $L$, the magnetic field
strength on the star surface $B_{0}$, the mass inside the flux rope $m_{0}$,
and the rate of magnetic reconnection $M_{A}$. Among them, the value of $M_{NS}$ is
fixed, the scale $L$ is taken as the radius of the neutral star
that is fixed, the magnetic field $B_{0}$ on the surface is also fixed,
and only $m_{0}$ and $M_{A}$ leave some room to be adjusted in our
calculations in order to obtain theoretical results that are in agreement with observations.
Below we shall see that even $m_{0}$ could be determined to a certain degree according to
observations.

As for $M_{A}$, unfortunately, there is not an acceptable value or theory for its value in the case
of the eruption on the magnetar for the time being. So we have to resort to the existing results
for $M_{A}$ in the case of solar eruptions. Basically, the standard theory of magnetic reconnection
requires that $M_{A}$ range from 0 to 1 \citep[e.g., see ][]{PF00,Lel03}, and
observations of solar eruptions put the value of $M_{A}$ in the range from 0.001 to 0.1
\citep[e.g., see ][]{Yok01,Lel05,Lel07}. This gives us some flexibility to choose the value of $M_{A}$
so that our theoretical results would be consistent with observations. We understand that this
inevitably brings unexpected uncertainties to our results, and we look forward to more new
observations that will help us constrain these free parameters and improve our model further in
the future.

Since the energy involved in the
magnetar giant flare requires that a large fraction of the magnetosphere be affected,
the typical size of an active region is of the order of the neutron star radius \citep{Lyu06}.
The distance between the two point-sources, $2\lambda_{0}$, and the length of the
flux rope, $L=2\lambda_{0}$, are of the order of the neutron star radius, $R_{0}$;
and we choose $\lambda_{0}= R_{0}/2$ in our calculation because this scale of the magnetic field
may provide enough magnetic energy to power a giant flare.
The Australia Telescope Compact Array and the Very Large Array observed an expanding radio nebula
associated with the giant flare from magnetar SGR 1806- 20 \citep{Tay05,Cam05,Gae05},
which implies a large scale mass ejection in the eruptive process that produced the giant flare.
Similar explosive phenomenon of the giant flare together with a mass ejection may also occur in
a black hole accretion disk system, which is also known as an episodic jet \citep[e.g., see also][]{Yel09}.
If the ejecta is roughly spherical, the radiation is well-explained by synchrotron
emission from a shocked baryonic shell with a mass of $\geq 10^{24.5}$ g and an
expansion velocity of $\simeq 0.4c$ \citep{Gae05,Gra06,Mas10}. We choose $m = 10^{26}$~g
because the ejecta can be ejected outward successfully at a reasonable velocity. If
the mass is too heavy, it may not be thrown away successfully; on the other hand,
if the ejecta is too light, it will travel too fast to fit the observational result.
According to these facts,
we choose the characteristic values of these parameters mentioned above as below:

$R_{0} = 10^{6}$ cm, $\lambda_{0} = 5 \times 10^{5}$ cm, $ m = 10^{26}$ g.

Solving equation (\ref{dht}) through (\ref{d2ht2}) with $M_{A} = 0.02$
(according to our experience in dealing with the solar case) gives
the time profiles of the flux rope height $h$, the lower top and the upper top,
$p$ and $q$, of the current sheet, respectively. The results are plotted in
Figure \ref{Fig.3}, and the corresponding velocity of the flux rope is plotted
in Figure \ref{Fig.4}, which indicates that the flux rope is thrust away from
the star surface very quickly and can be easily accelerated to a speed of $0.5c$
within several milliseconds following the catastrophe. This time interval roughly
fits that of the impulsive phase of the eruption from SGR 1806-20. We shall
further discuss the issue related to the time scale of the eruption later.
The released magnetic energy is enough to be converted into kinetic energy
that could accelerate the flux rope to high speed. The flux rope is thrust
outward by the catastrophe rapidly at the beginning stage of the eruption,
and the reconnection process invoked in the current sheet following the catastrophe
allows the flux rope to propagate continuously at high speed for a while
\citep[see discussions of][]{LF00,Lin02}. Therefore, the flux
rope or the ejecta could go very far away from the central star after the
eruption commences.

Then we get the variations
of the total magnetic flux sent into interstellar space by reconnection versus
time for magnetar SGR 1806- 20 giant flare for $B=10^{15}$ G in Figure \ref{Fig.5a}.
The preexisting magnetic flux inside the flux rope prior to the eruption
was $\Phi_{p}= I_{0}\pi/c L\approx7.85\times10^{26}$~Mx \citep[see also][]{Lel04,LS04},
and the total magnetic flux brought into space by reconnection in the whole process
is about $6.7\times 10^{26}$~Mx, where Mx $= 10^{-8}$~Wb is the CGS unit of the magnetic flux .
Therefore, the eruption brought more than
$1.5\times 10^{27}$~Mx of magnetic flux away from the central star.
Time histories of the total magnetic flux of SGR 1900+14
and SGR 0526-66 related to reconnection are also shown in Figure \ref{Fig.5a}
for $B=5.0\times10^{14}$~G
for both events, and the total magnetic fluxes sent into interstellar space during the
eruption were $6.9\times 10^{26}$~Mx for SGR 1900+14 and $6.7\times 10^{26}$~Mx for
SGR 0526-66, respectively.

The energy budget for the eruption in our model is calculated and shown in Figure \ref{Fig.5}.
The total energy involved in the eruption $W_{0}$ consists of free magnetic energy,
kinetic energy, gravitational potential energy and radiative energy (see also equation (\ref{wtal})).
If we take the gravitational potential energy to be zero right before the catastrophe,
$W_{0}$ is identified with $W_{mag}$ as the catastrophe is invoked. In the process of
eruption $W_{mag}$ keeps decreasing and the other types of energy as the left side of
equation (\ref{wtal}) continuously increase.

Our calculations indicate that the free magnetic energy stored in the system
prior to the eruption is more than $10^{47}$ erg which seems to be enough to power the
giant flare from magnetar SGR 1806 - 20 on 27 December 2004. But we need to check
whether all of this stored magnetic energy or just part of it could be quickly converted into
radiation and accelerating plasma to account for the giant flare and the associated
mass ejection. Usually,
the eruption starts with the catastrophe and reconnection helps the eruption develop
smoothly \citep{LF00,Lin02}. So the rate of the energy release is governed
by both the catastrophe and the following reconnection process. Generally, the energy
release rate reaches maximum in the catastrophic stage. But two facts prevent most of
the energy release from taking place in the catastrophic stage: first, the time scale
of the catastrophe is too short (several hundred seconds for a solar eruptive event and
several tenth ms for the magnetar case) to allow the main energy release to occur; and
second, the ideal MHD nature of the catastrophe leads to the development of a current
sheet that halts the further evolution after a very short period in the initial stage
unless the ideal MHD environment breaks down \citep[e.g., see][]{FI91,Iea93}.

Compared to the catastrophe, the reconnection has long time scale (a few 10$^{3}$
seconds in the solar case, and a few ms in the magnetar case), so the energy conversion
could be taking place gradually and lasting long. Additionally, reconnection is a totally
non-ideal MHD process, and the magnetic field can be quickly diffused with its energy being
converted into other types of energy during this process, thus allowing the eruptive process to develop
smoothly.

Therefore, in an eruptive process, no matter whether on the Sun or on a magnetar,
 most of the free energy is converted into radiative and kinetic energies via magnetic
reconnection \citep[see detailed discussions on this issue by][]{FPI94,FP95}, and the
time scale of the magnetic reconnection
determines the subsequent evolution in the system. Figure \ref{Fig.5} plots time
 profiles of various types of energy for SGR 1806-20, from which we see that after a transient
 stage of a few tenth ms (see also Figure \ref{Fig.4}), the evolution in the system turns to
 a gradual phase within several ms, and the main energy conversion happens within this several ms.

In addition, we also noticed that the energy conversion lasts for a very long period at a very
low rate (but not zero) after the progress discussed above, and that not all the free energy
is released in the whole process although $W_{mag}$ drops apparently from its initial value
to a low value. The difference
 between the initial value of $W_{mag}$ and that of $W_{mag}$ at least 500 seconds after is more than
$1 \times 10^{47}$ erg, which is enough to produce a giant flare
and the associated mass ejection that have ever been observed by several instruments.

After the catastrophe takes place, the flux rope is quickly ejected away from the star's
surface, a small fraction of released magnetic energy is partly converted into kinetic
energy and is partly used to do the work against the gravity in the catastrophe
(see also Figure \ref{Fig.2}). Formation of the X-point is followed by the
development of the current sheet, which is expected to be long because the reconnection
process is slow compared to the catastrophe. The existence of the X-point and / or the
current sheet allows reconnection to occur, and the magnetic energy is converted into
radiative and kinetic energy. In this process, the magnetic field at either side of current
sheet, which is responsible for the magnetic tension that prevents the flux rope from
moving far, approaches the shear giving rise to the Poynting flux into the sheet. Magnetic
reconnection dissipates this magnetic field, weakening the magnetic tension, heating the
plasma, and allowing the flux rope to escape from the center star \citep[see also discussions in
detail by][]{FL00,LF00,Lin02}.

As shown in Figure \ref{Fig.5}, $W_{mag}$ quickly decreases in the early stage, and the
kinetic and gravitational potential energy of the flux rope, together with the Poynting
flux and the associated emission, increases rapidly. This process occurs within the first millisecond
in a very energetic fashion. When the flux rope achieves escape velocity, the energy
released approaches an asymptotic value at long times. It is simple and straightforward
to prove that the amount of decrease in $W_{mag}$ is just equal to the sum of increases
in $W_{ke}$, $W_{gra}$ and $W_{EM}$.

The power output associated with radiative energy is directly related to
$S(t)$ as indicated by equation (\ref{st}) and the related discussions right after the
equation. So the time profile of $S(t)$ could be
considered the light curve of the event. Figure \ref{Fig.6} gives $S(t)$ as a function
of $t$. It shows that after the neutral point forms, the energy output rate curve has
a hard spike in the first several milliseconds, and then turns to a tail emission lasting
hundreds seconds. The shape of the light curve given by our model is very reminiscent of
the light curve of the giant flare from magnetar SGR 1806- 20 on 27 December 2004.
To compare our results with those of the giant flare from magnetar SGR 1806- 20 on
27 December 2004 observed by RHESSI $\gamma$- ray detectors, we create a composite
of the modified light curve from Figure \ref{Fig.6} and the observed one (see
Figure \ref{Fig.7}).

As shown in Figure \ref{Fig.7}, the black curve is from the giant flare 20-100-keV
time history plotted with 0.5-s resolution, and red curve is from our model.
The time $t=0$ corresponds to 77,400 s UT. In this plot, the flare began
with the spike at 26.64 s and saturated the detectors within 1 ms. The detectors emerged
from saturation on the falling edge 200 ms later and remained unsaturated after that.
Photons with energies $\geq 20 $ keV are unattenuated; thus the amplitude variations
in the oscillatory phase are real, and are not caused by any known instrumental effect
\citep{Hur05}. We noticed in Figure \ref{Fig.7} that the observed data consist of 3
phases: a ~ 1-s long precursor, an initial spike of the giant flare which lasts for ~ 0.2~s
and an oscillatory tail modulated with a period of 7.56 s.

On the basis of our model and with assumptions that the energy was composed of photons
with an average energy of 60 keV and that the detected energy is proportional to
$\sim 4\pi d^{2} S(t)$, where $d \approx 15$  kpc is the distance between the Earth and the SGR 1806- 20
\citep{CE04,Cam05,MG05}, we obtain the light curve of the giant flare for the case where
the outburst is spherical for simplicity.
The time when the X-point appears
corresponds to 77,426.6~s UT on 27 December 2004 in our calculations. In Figure \ref{Fig.7},
We see a good agreement of our model with observations.
In this work, we do not include the precursor of the giant flare because our calculations
start at the moment when the system loses its equilibrium. The precursor may be explained
by the plastic deformation of the crust \citep{Lyu06}, crust motions \citep{Rud91a}, or
other mini scale eruptions like what happens frequently in the solar eruptions \citep[e.g., see][]{Jel11,SLS12}.

Using the same methods, we are also able to calculate the light curves of the giant flares from
SGR 0526-66 and SGR 1900+14,
and compare them with the observations as shown in Figures \ref{Fig.8} and \ref{Fig.9},
respectively. The observational data for SGR 0526-66 were detected by Venera 11 \citep{Maz79},
and the energy range covered 50-150 keV, the time cadence of the observation was 1/4~s,
and the burst onset time $t_{0}$ for
Venera 11 was 15~h 51~min 39~s, 145~UT. The distance used in our calculation was
$d = 55 $ kpc \citep{Maz79} for this event.

The observations for
the 27 August 1998 event were performed by Ulysses from the 0.5-s resolution continuously
available real-time data \citep{Hur99}, which had 25-150 keV time
history and were corrected for dead-time effects. Zero seconds corresponds to 37,283.12s
UT at the Earth, and $d = 7 $ kpc \citep{Hur99} for SGR 1900+14. As shown in
Figures \ref{Fig.8} and \ref{Fig.9}, our results fit the observational ones well.

\section{Discussions and Conclusions}
We develop a theoretical model via an analytic approach  for a magnetar giant flare in the
framework of the solar CME catastrophe model \citep{LF00}. We considered the physical
process that causes the catastrophic loss of equilibrium
of a twisted flux rope in the magnetar magnetosphere.
The model was constructed according to manifestations of the eruption from the
 magnetar SGR 1806-20. As happens on the Sun, the free magnetic energy that drives
 the eruption on the magnetar is slowly stored in
the magnetosphere of the magnetar long before the outburst until the system loses
 mechanical equilibrium, and then is quickly released in the consequent eruptive process,
also known as a magnetar giant flare. In our model, the energy driving the eruption comes from the magnetosphere.
The motion of footpoints causes the magnetic configuration to
lose its equilibrium and release magnetic energy eventually, which is in principle the same as the
results of numerical simulations \citep{Pfr12a,Pfr12b,Pfr13,Yu11,Yu12}.

In the present model, the eruption starts with a loss of equilibrium in the magnetic configuration,
which could be triggered either by the change in the background magnetic field, or by the
sudden break of a crust piece on the surface of the star where the disrupting magnetic
configuration roots in. The evolution in this stage could be purely mechanical, namely no
dissipation or magnetic reconnection needs to take place in the magnetosphere. But the
dissipation is required in the consequent progress, otherwise the conversion of magnetic
energy into radiative and kinetic energy would be stopped, and the evolution in the system
ceases without plausible eruptive phenomenon happening \citep[e.g., see detailed discussions by][]{
FL00,LF00,Lin02,PF02,Lel03,Fel06}.

Our main results deduced from the present analysis are summarized as follow:

1. The system could store free energy of
more than $10^{47}$~ergs prior to the eruption, which is enough to drive a giant flare like
the one from magnetar SGR 1806-20  observed on 27 December 2004.

 2. A combination of the following three factors determines that the disrupting magnetic
 configuration is highly stretched, and a long magnetically neutral current sheet forms
 separating two magnetic fields of opposite polarity. These factors are the mechanical
 property of the loss of equilibrium, the inertia  of the magnetic field that
 tends to keep the original topological  features in the configuration unchanged
 without diffusion in the system \citep[see detailed discussions by][]{FI91,Iea93}, and the fact
 that the reconnection time-scale is long compared to that of the loss of equilibrium.

3. A long current sheet is usually unstable to perturbations due to various plasma instabilities,
such as the tearing mode instability, and the consequent turbulence quickly dissipates the magnetic
field such that the stored magnetic energy is rapidly converted into radiative and kinetic energy
of the plasma inside the current sheet, as well as the kinetic energy and
the gravitational potential energy of the mass in the ejected flux rope. Most of the released
magnetic energy turns into radiative energy in the case of the giant flare from SRG 1806-20.

4. Conversion of energy by
reconnection mainly occurs at the lower part of the current sheet, and the hottest part of
the current sheet is then expected close to the star surface. A large amount of energetic
particles and a heat conduction front are also created by reconnection and propagate downward
along magnetic field lines. They eventually reach the star surface and may heat the relevant
region significantly, which then leads to the formation of a ``fireball" near the
neutron star's surface. Therefore, a ``fireball" is a straightforward and natural consequence
of our model.

5. We calculated the light curve on the basis of our model
and compare it with the observational one obtained by the RHESSI $\gamma$- ray detectors.
We note that the calculated light curve consists of a hard spike lasting a half ms and a tail
emission last a few $10^{2}$~s, which are consistent with the observations.

6. Our calculations indicate that the magnetic flux preexisting in the flux rope before the
eruption was about $7.9\times 10^{26}$~Mx, and that the magnetic flux brought from the environment
around the disrupting magnetic field into the ejecta bubble was around $6.8\times 10^{26}$~Mx.
Therefore, total magnetic flux of more than $1.5\times 10^{27}$~Mx was sent into interstellar
space from the central star by the super eruption from SGR 1806-20.

7. We duplicated our calculations for the giant flares from SGR 0526-66 and SGR 1900+14,
respectively, and also found good agreement of our model with observations. Furthermore, the
the total magnetic fluxes ejected into interstellar space by the eruption
were $6.9\times 10^{26}$~Mx for SGR 1900+14 and $6.7\times 10^{26}$~Mx for
SGR 0526-66, respectively.

\acknowledgments
We are grateful to K. Hurley for providing the observational data used in this paper,
and to Z. Dai and Y. Yuan for valuable discussions and suggestions.
Helpful advice and suggestions for using the observational data given by
D. M. Palmer are highly appreciated. We appreciate the referee for constructive advice
and comments for improving this work as well. This work was supported by
Program 973 grants 2011CB811403 and 2013CBA01503, NSFC grants 11273055
and 11333007, and CAS grants KJCX2-EW-T07 and XDB09000000. F. Y. was supported
by the NSFC grants 11121062 and 11133005. K. K. R. is supported by NSF grant AGS-1156076
to the Smithsonian Astrophysical Observatory.

\appendix
\section{Appendix}
The motion of flux rope is described by four parameters ($h$, $p$, $q$, $\dot{h}$),
which are the height of flux rope, heights of the lower and higher tips of the
current sheet, and the velocity of flux rope, respectively.
Another three independent equations are required in solving Equation (\ref{Fmg}).
In this section, the required equations are simply described. They are based
on \citet{LF00}, \citet{LMV06} and \citet{Yel09}.

\begin{figure}[fig1]
\epsscale{0.9}
\plotone{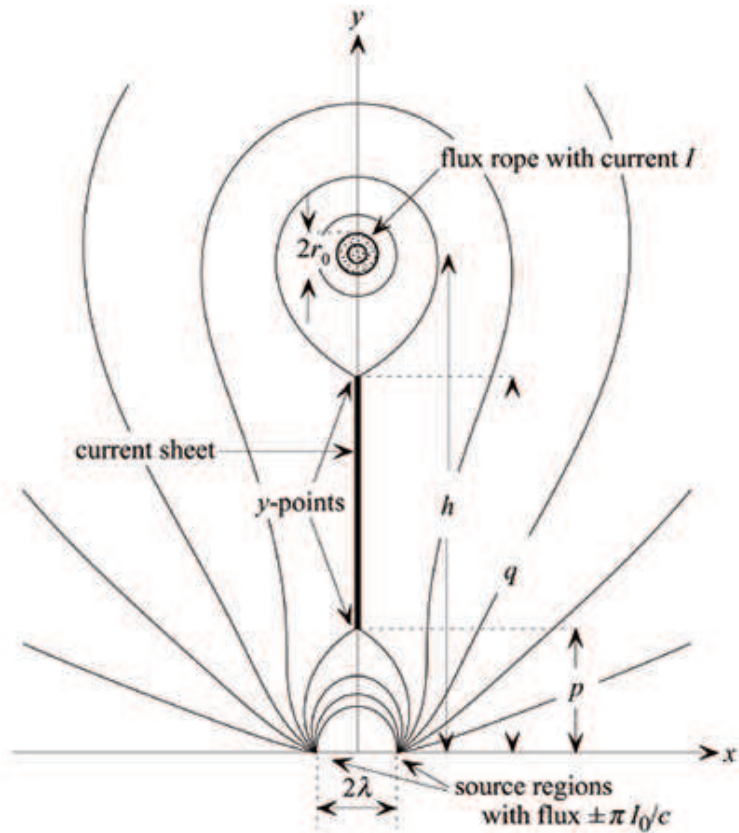}
\caption{Diagram of the flux
rope configuration, showing the mathematical notation used in the text
from \citep{LF00}. The $x$-axis is located on the star surface, and the
$y$-axis points upward. The height of the center of the flux rope is
donated by $h$, $p$, and $q$ denote the lower and upper tips of
the current sheet, respectively, and the distance between the magnetic
source regions on the photosphere is 2$\lambda.$ }\label{Fig.1}
\end{figure}

\begin{figure}[fig2]
\epsscale{1}
\plotone{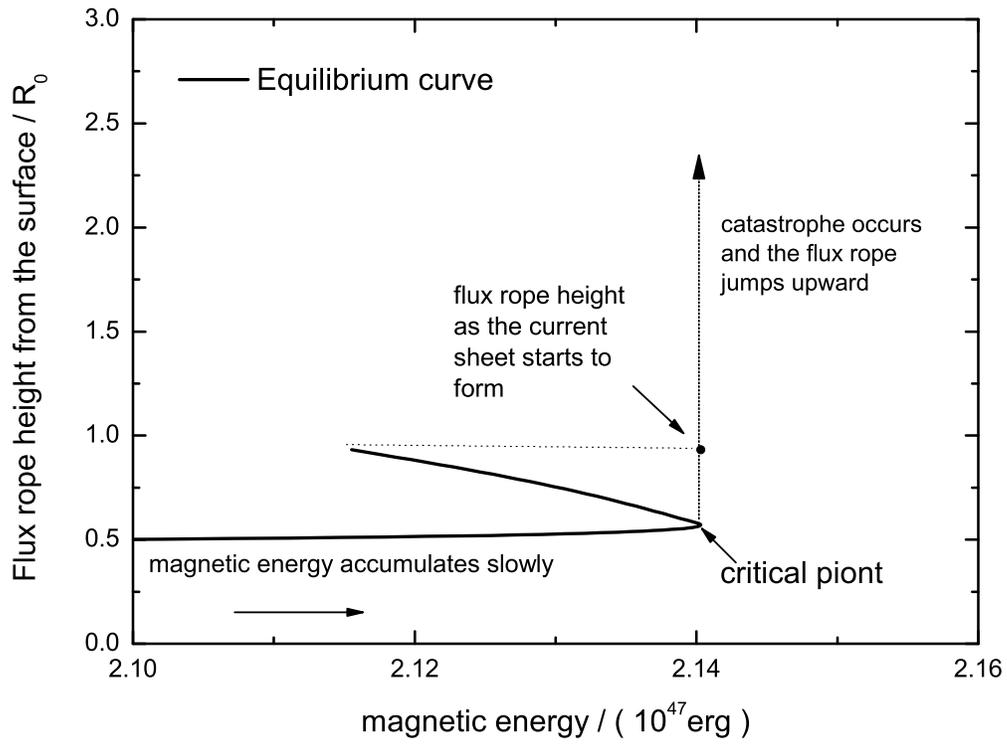}
\caption{Equilibrium flux rope height (in units of the neutron star radius
$R_{0} \approx 10^{6}$ cm) as a function of stored magnetic energy.
When the critical point is reached, the catastrophe occurs,
and the flux rope is ejected outwards.}\label{Fig.2}
\end{figure}

\begin{figure}[fig3]
\epsscale{1}
\plotone{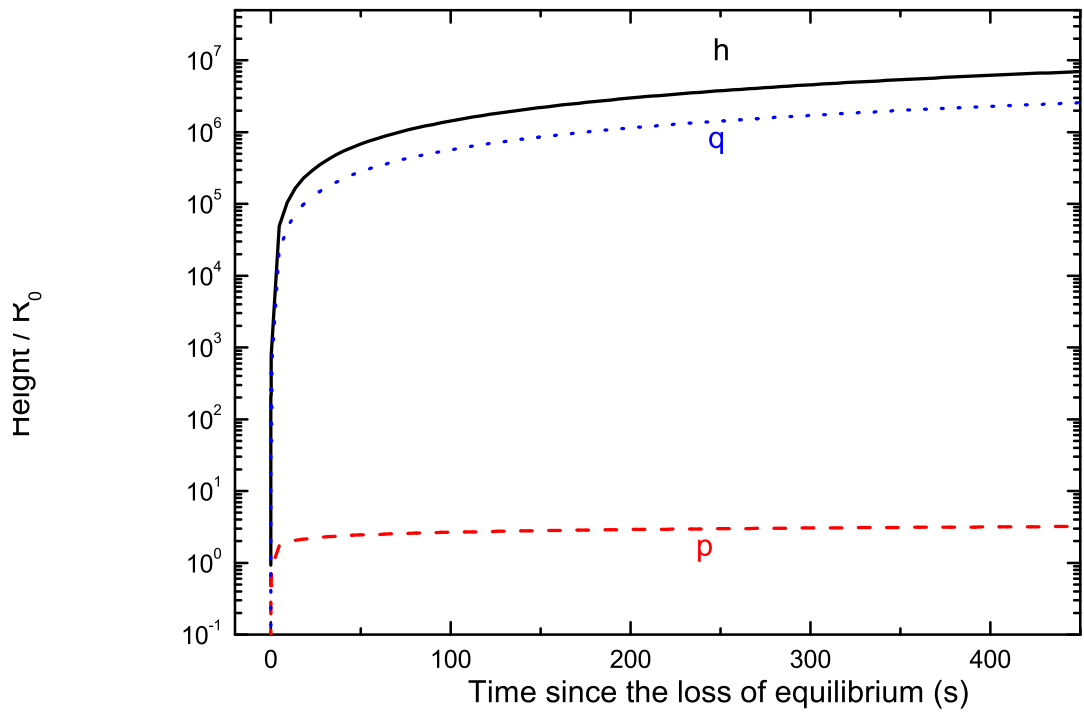}
\caption{Plots of the height of the flux rope, $h$, and the top and the bottom tips of the
current sheet ($q$, $p$) as function of time.}\label{Fig.3}
\end{figure}

\begin{figure}[fig4]
\epsscale{1}
\plotone{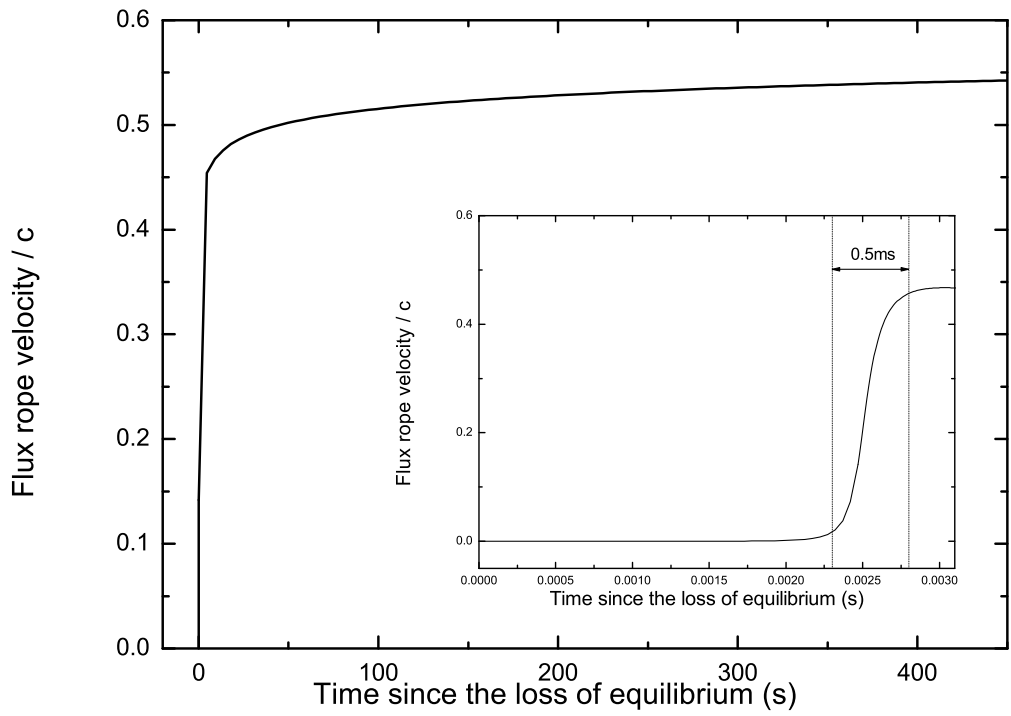}
\caption{Plots of the velocity of the flux rope as a function of time. The inset
describes the detailed evolution in the speed of the flux rope, which implies a
very energetic eruption.}\label{Fig.4}
\end{figure}

\begin{figure}[fig5a]
\epsscale{1}
\plotone{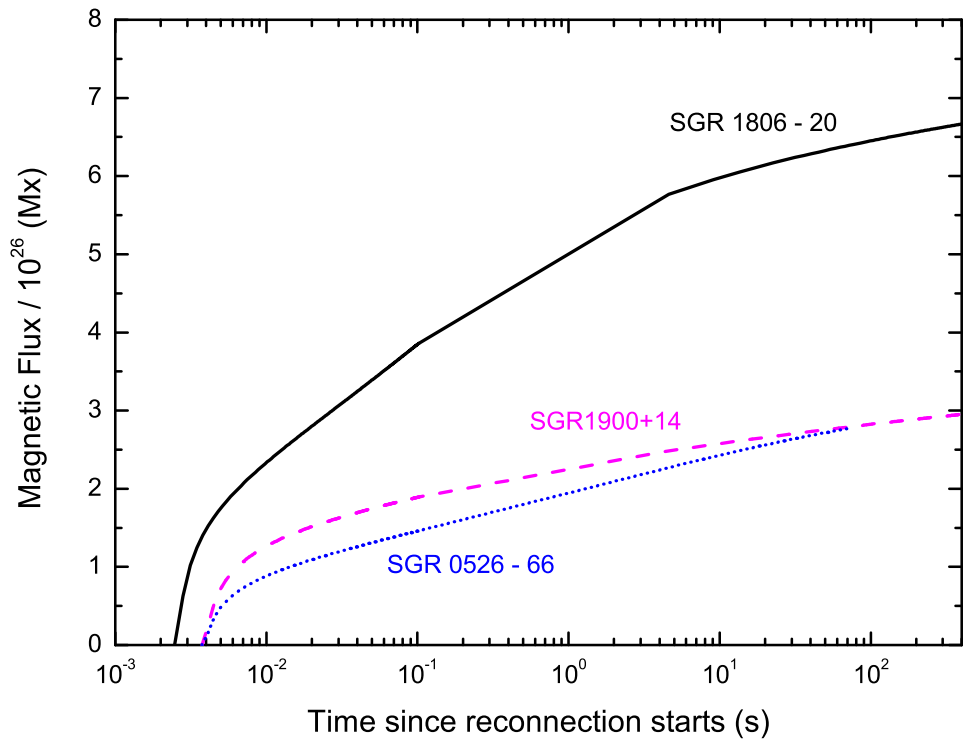}
\caption{Variations of the total magnetic flux sent into interstellar space
by reconnection versus time for three giant flares occurring on different
magnetars.}\label{Fig.5a}
\end{figure}

\begin{figure}[fig5]
\epsscale{1}
\plotone{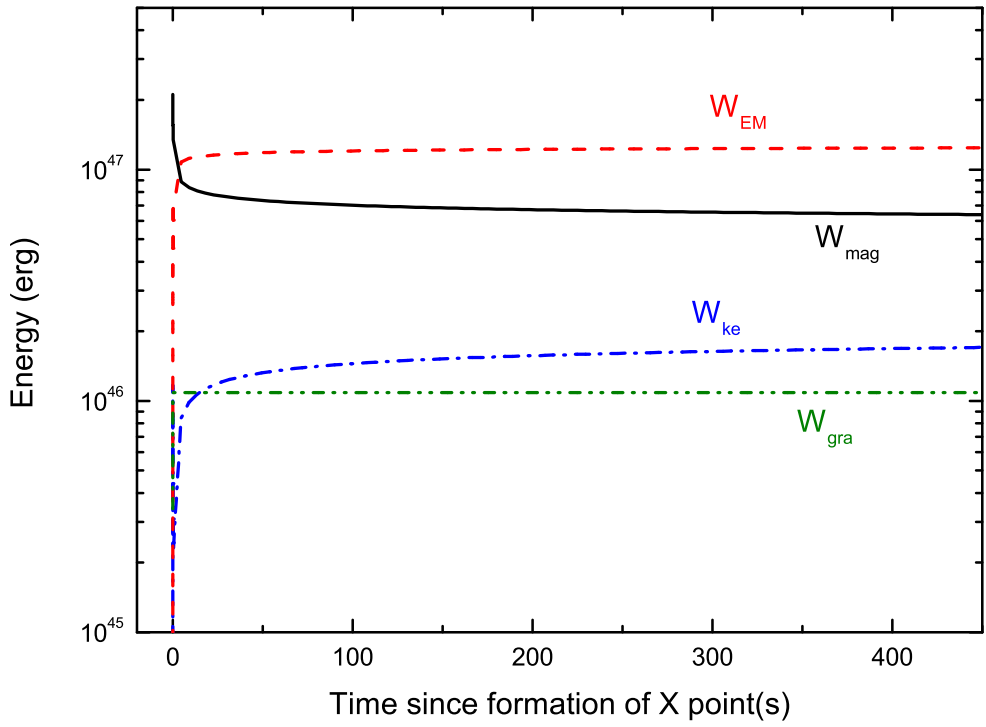}
\caption{Plots of magnetic energy $W_{mag}$, radiative energy $W_{EM}$, gravitational energy $W_{gra}$
 and kinetic energy $W_{ke}$ as function of time for SGR 1806-20.}\label{Fig.5}
\end{figure}

\begin{figure}[fig6]
\epsscale{1}
\plotone{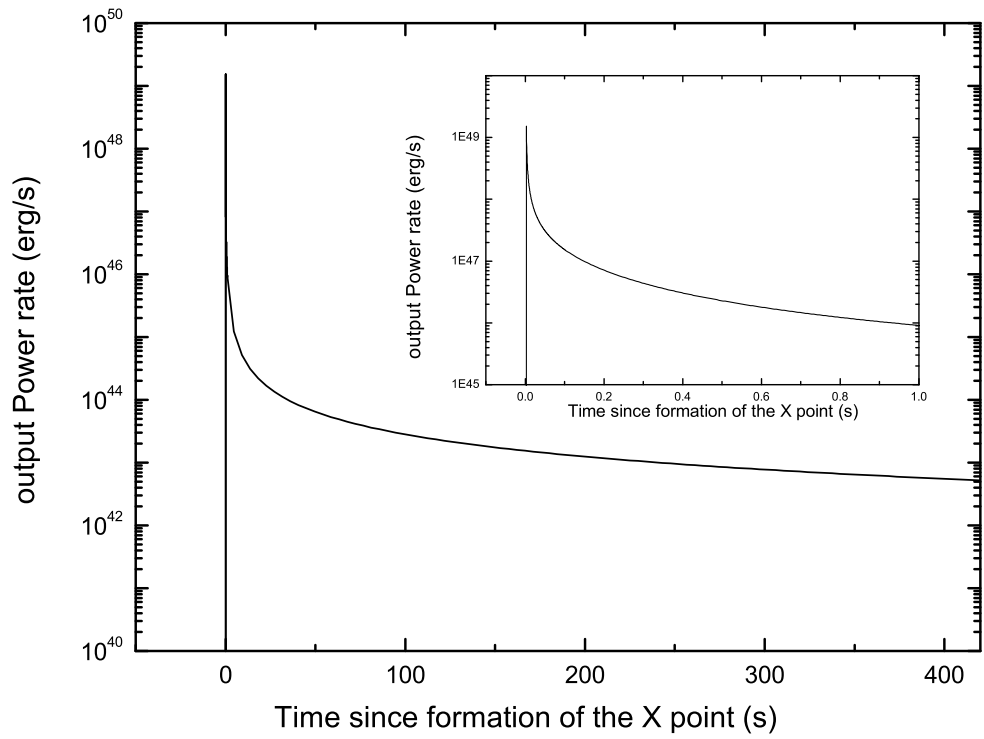}
\caption{Light curve of the eruption governed by $S(t)$ given in equation
(\ref{Sst}). }\label{Fig.6}
\end{figure}

\begin{figure}[fig7]
\epsscale{1}
\plotone{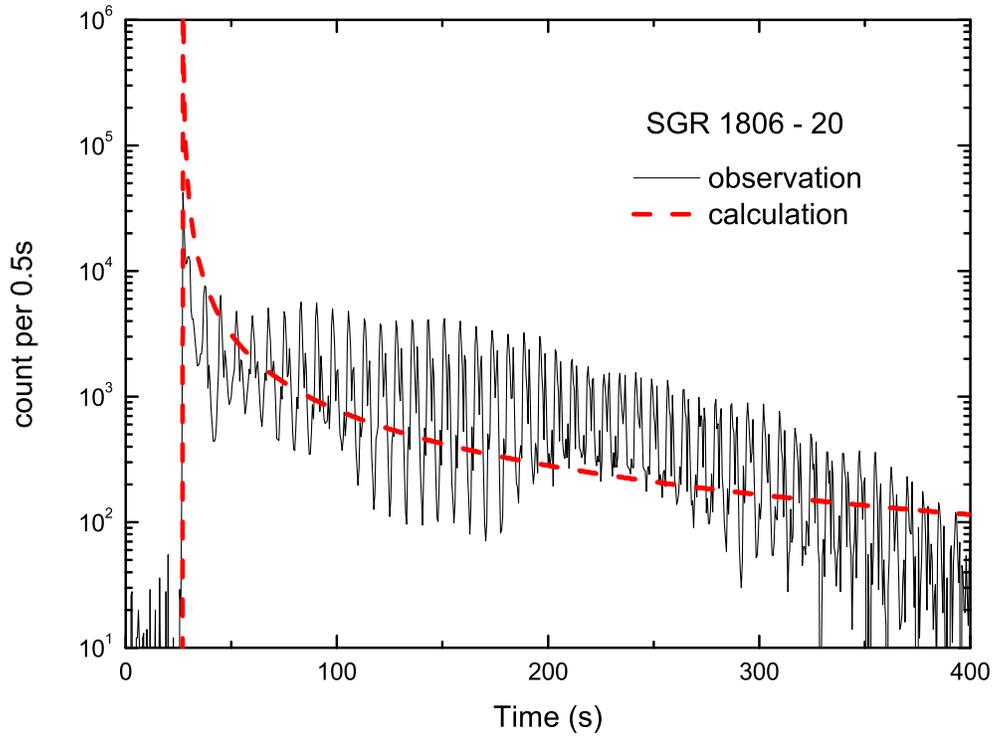}
\caption{Light curve for the SGR 1806-20 giant flare given by calculations
compared with the observations from the RHESSI spacecraft. }\label{Fig.7}
\end{figure}

\begin{figure}[fig8]
\epsscale{1}
\plotone{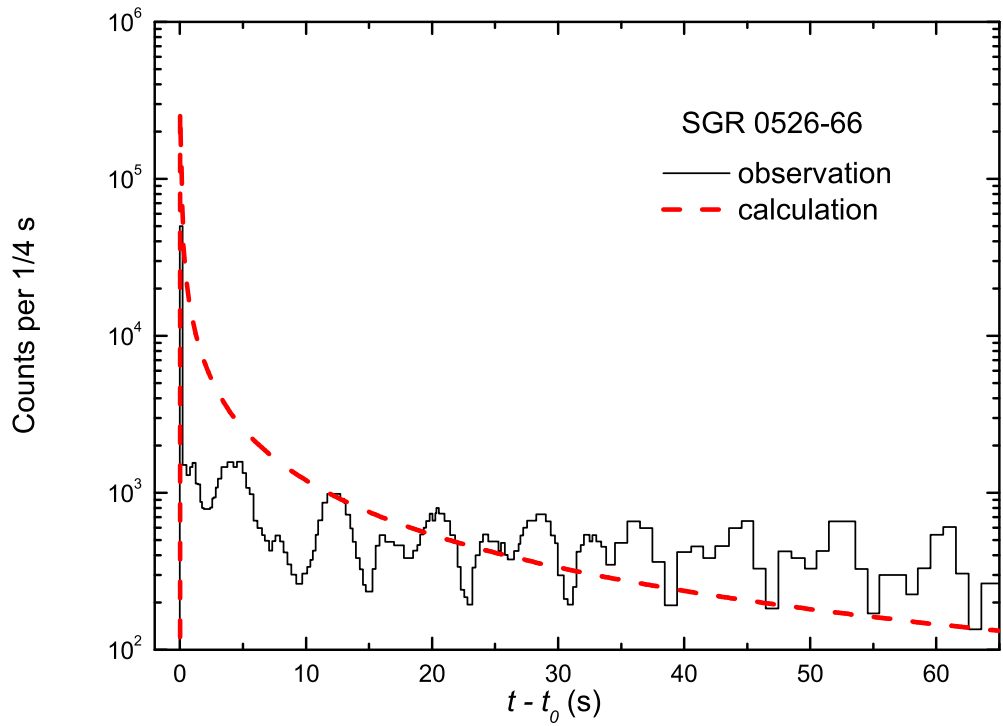}
\caption{Light curve for the SGR 0526-66 given by calculations compared
with the observations from the Venera spacecraft. }\label{Fig.8}
\end{figure}

\begin{figure}[fig9]
\epsscale{1}
\plotone{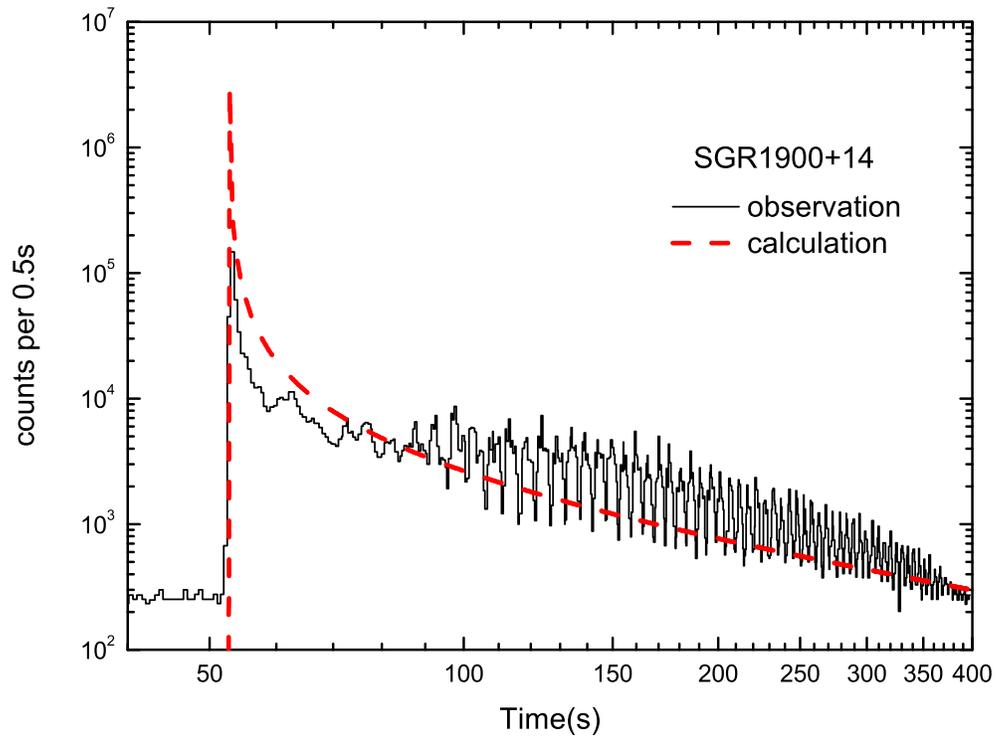}
\caption{Light curve for the SGR 1900+14 given by calculations compared
with the observations from the Ulysses spacecraft. }\label{Fig.9}
\end{figure}

The forces acting on the flux rope include the magnetic force, $F_{m}$ (the first term on
the right hand side of equation (\ref{Fmg}), and the gravity, $F_{g}$ (the second term), which
read as
\begin{eqnarray}
{F_m} = \frac{{{B_0}^2{\lambda ^4}}}{{8h{L_{PQ}}^2}}\left[\frac{{{H_{PQ}}^2}}{{2{h^2}}} - \frac{{({p^2}
 + {\lambda ^2})({h^2} - {q^2}) + ({q^2} + {\lambda ^2})({h^2} - {p^2})}}{{{h^2} + {\lambda ^2}}}\right],
\label{a1}
\end{eqnarray}%

\begin{eqnarray}
{F_g} = \frac{{G{M_{NS}}\gamma{m_0}}}{{{{\left( {{R_0} + h} \right)}^2}}},
\label{a2}
\end{eqnarray}%
respectively, where $L_{PQ}^{2}=(\lambda^{2}+p^{2})(\lambda^{2}+q^{2})$,
$H_{PQ}^{2}=(h^{2}-p^{2})(h^{2}-q^{2})$,
$B_{0}$ and $\lambda$ are the surface magnetic field strength and the half
distance between the two point sources on the surface of the star prior to
the eruption, respectively (see also Figure \ref{Fig.1}). On the right-hand
side of equation (\ref{a1}), the two terms in the square brackets
denote the magnetic compression force and the magnetic tension.
The magnetic compression force pushes the flux rope upwards
and the magnetic tension pulls the flux rope downwards. The system
is in equilibrium as they balance each other. When the system loses its
equilibrium, the compression dominates the tension, thrusting the flux
rope outward in a catastrophic way.

The force-free condition outside the current sheet and the frozen
flux condition on the surface of the flux rope can be used to deduce
the equations governing the evolution of the current sheet, which are
\citep{FI91,LF00}
\begin{eqnarray}
\frac{{dp}}{{dt}} = p' \dot{h}, \nonumber \\
\frac{{dq}}{{dt}} = q' \dot{h},
\end{eqnarray}%
where $p'$ and $q'$ are given in equations (\ref{dpdh})
and (\ref{dqdh}) with
\begin{eqnarray}
\tilde{A_{0h}}= \frac{c E_{z}}{ \dot{h}}+ A_{0h}
= \frac{M_{A}B_{y}(0,y_{0})c}{ \dot{h}}+ A_{0h},
\end{eqnarray}
where $B_{y}(0,y_{0})$
can be deduced from equation (\ref{Bxy}), and
\begin{eqnarray}
{A_R} &= &\frac{{\lambda {H_{PQ}}}}{{2h{L_{PQ}}}}
\ln \left[ \frac{{\lambda {H_{PQ}}^3}}{{{r_{00}}{L_{PQ}}({h^4} - {p^2}{q^2})}} \right]  \nonumber \\
  &+& \tan^{-1}\left( \frac{\lambda }{h}\sqrt {\frac{{{p^2} + {\lambda ^2}}}{{{q^2} + {\lambda ^2}}}} \sqrt {\frac{{{h^2} - {q^2}}}{{{h^2} - {p^2}}}} \right)
  + \frac{\lambda }{{q{L_{PQ}}}} \left\{ ({h^2} - {q^2})F\left[ {\sin^{-1}}\left(\frac{q}{h}\right),\frac{p}{q} \right] \right. \nonumber \\
  &+& \left. ({q^2} - {p^2})\Pi \left[ {\sin^{-1}}\left(\frac{q}{h}\right),\frac{{{p^2} + {\lambda ^2}}}{{{q^2} + {\lambda ^2}}},\frac{p}{q} \right] - \frac{{{H_{PQ}}^2}}{{{h^2}}}\Pi \left[ {\sin^{-1}}\left(\frac{q}{h}\right),\frac{{{p^2}}}{{{h^2}}},\frac{p}{q} \right]  \right\} \nonumber \\
  &=& \frac{\pi }{4} + \ln \left( \frac{{2\lambda }}{{{r_{00}}}} \right), \nonumber \\
 {A_{Rp}} &= & \frac{{\lambda p({h^2} + {\lambda ^2})}}{{q{{({p^2} + {\lambda ^2})}^2}}}\sqrt {\frac{{{p^2} + {\lambda ^2}}}{{{q^2} + {\lambda ^2}}}} \left\langle \left(1 - \frac{{{p^2}}}{{{h^2}}} \right)
 \Pi \left[{\sin^{-1}}\left(\frac{q}{h}\right),\frac{{{p^2}}}{{{h^2}}},\frac{p}{q} \right] \right. \nonumber \\
  &-& F \left[ {\sin^{-1}}\left(\frac{q}{h}\right),\frac{p}{q} \right]
  - \left. \frac{q}{{2h}}\sqrt {\frac{{{h^2} - {q^2}}}{{{h^2} - {p^2}}}} \left\{1 + \ln \left[ \frac{{\lambda {H_{PQ}}^3}}{{{r_{00}}{L_{PQ}}({h^4} - {p^2}{q^2})}}\right] \right\} \right\rangle, \nonumber \\
 {A_{Rq}} &=& \frac{{\lambda ({h^2} + {\lambda ^2})}}{{{{({q^2} + {\lambda ^2})}^2}}}\sqrt {\frac{{{q^2} + {\lambda ^2}}}{{{p^2} + {\lambda ^2}}}} \left\langle \left(1 - \frac{{{p^2}}}{{{h^2}}} \right) \right.
 \Pi \left[ {\sin^{-1}}\left(\frac{q}{h}\right),\frac{{{p^2}}}{{{h^2}}},\frac{p}{q} \right] \nonumber\\
 &-& \left. F\left[ {\sin^{-1}}\left(\frac{q}{h}\right),\frac{p}{q} \right]
  - \frac{q}{{2h}}\sqrt {\frac{{{h^2} - {p^2}}}{{{h^2} - {q^2}}}} \left\{1 + \ln \left[\frac{{\lambda {H_{PQ}}^3}}{{{r_{00}}{L_{PQ}}({h^4} - {p^2}{q^2})}}\right] \right\} \right\rangle, \nonumber \\
 {A_{Rh}} &= &\frac{\lambda }{{2{h^2}{L_{PQ}}{H_{PQ}}}}\left\{ 2\frac{{{h^6} - {\lambda ^2}{p^2}{q^2}}}{{{h^2} + {\lambda ^2}}} - \frac{{{h^2}({p^2} + {q^2})({h^2} - {\lambda ^2})}}{{{h^2} + {\lambda ^2}}} \right.\nonumber \\
  &+& ({h^4} - {p^2}{q^2})\left. \ln\left[ \frac{{\lambda {H_{PQ}}^3}}{{{r_{00}}{L_{PQ}}({h^4} - {p^2}{q^2})}} \right] \right\}  + \frac{\lambda }{{hq{L_{PQ}}}} \nonumber \\
  &\times& \left\{ ({h^2} + {q^2})F\left[{\sin^{-1}}\left(\frac{q}{h}\right),\frac{p}{q}\right] - {q^2}E\left[{\sin^{-1}}\left(\frac{q}{h}\right),\frac{p}{q}\right] \right. \nonumber \\
  &-&\left. \frac{{{h^4} - {p^2}{q^2}}}{{{h^2}}}\Pi \left[ {\sin^{-1}}\left(\frac{q}{h}\right),\frac{{{p^2}}}{{{h^2}}},\frac{p}{q}\right] \right\}.
\label{a4}
\end{eqnarray}
and
\begin{eqnarray}
 A_{0}^{0} &=& \frac{{2{I_0}}}{c}\frac{\lambda }{{q{L_{PQ}}}}
 \left[ ({h^2} - {q^2})K\left(\frac{p}{q}\right) + ({q^2} - {p^2}) \right. \nonumber \\
  &\times & \Pi \left(\frac{{{p^2} + {\lambda ^2}}}{{{q^2} + {\lambda ^2}}},\frac{p}{q}\right) - \left. \frac{{{H_{PQ}}^2}}{{{h^2}}}\Pi \left(\frac{{{p^2}}}{{{h^2}}},\frac{p}{q} \right) \right], \nonumber\\
 {A_{0p}}& = & \frac{{\lambda p({h^2} + {\lambda ^2})({q^2} + {\lambda ^2})}}{{{h^2}q{{[({p^2} + {\lambda ^2})({q^2} + {\lambda ^2})]}^{3/2}}}} \nonumber \\
  & \times & \left[ ({h^2} - {q^2})\Pi \left(\frac{{{p^2}}}{{{h^2}}},\frac{p}{q}\right) -
  {h^2}K\left(\frac{p}{q}\right) \right],\nonumber \\
 {A_{0q}}&= &\frac{{\lambda ({h^2} + {\lambda ^2})({p^2} + {\lambda ^2})}}{{{h^2}{{[({p^2} + {\lambda ^2})({q^2} + {\lambda ^2})]}^{3/2}}}} \nonumber \\
  &\times & \left[ ({h^2} - {q^2})\Pi \left(\frac{{{p^2}}}{{{h^2}}},\frac{p}{q}\right) -
  {h^2}K\left(\frac{p}{q}\right) \right], \nonumber\\
 {A_{0h}} &= & - \frac{\lambda }{{{h^3}q\sqrt {({p^2} + {\lambda ^2})({q^2} + {\lambda ^2})} }} \nonumber \\
  &\times & \left[ {h^2}{q^2}E\left(\frac{p}{q}\right) - {h^2}({h^2} + {q^2})K\left(\frac{p}{q}\right) \right. \nonumber \\
 & + & \left. ({h^4} - {p^2}{q^2})\Pi \left(\frac{{{p^2}}}{{{h^2}}},\frac{p}{q}\right) \right], \label{a3}
\end{eqnarray}
where the specification of $K$ and $\Pi$ can be found in section 2.2; $F$ and $E$
are first and second kinds of incomplete elliptic integrals, respectively.

\clearpage

\clearpage

\clearpage

\clearpage

\clearpage

\clearpage

\clearpage

\clearpage

\clearpage

\clearpage

\clearpage

\end{document}